\documentclass[api,prl,showpacs,letterpaper,twocolumn,groupedaddress,footinbib,amssymb]{revtex4}

\usepackage{graphicx}

\newcommand{\wta}{Ta$_x$W$_{1-x}$\,}
\newcommand{\wtax}{Ta$_{.42}$W$_{.58}$\,}
\newcommand{\cA}{$c(2\times2)$-AFM}
\newcommand{\pA}{$p(2\times1)$-AFM}

\begin{document}

% Use the \preprint command to place your local institutional report number in the upper righthand corner
% of the title page in preprint mode.  Multiple \preprint commands are allowed.
% Use the 'proprietresses' class option to override journal defaults to display numbers if necessary.

%\preprint{}

\title{Magnetic Phase Control in Monolayer Films by Substrate Tuning}

\author{P.\,Ferriani$^{1}$}
\email[corresp.\ author: ]{pferrian@physnet.uni-hamburg.de}
\author{I.\,Turek$^2$}
\author{S.\,Heinze$^1$}
\author{G.\,Bihlmayer$^3$}
\author{S.\,Bl{\"u}gel$^3$}

\affiliation{$^1$Institute of Applied Physics and Microstructure Research Center,
    University of Hamburg, Jungiusstrasse 11, 20355 Hamburg, Germany\\
$^2$Institute of Physics of Materials, Academy of Sciences of the Czech Republic, 
    Zizkova 22, 61662 Brno, Czech Republic\\
$^3$Institut f\"{u}r Festk\"{o}rperforschung,
    Forschungszentrum J\"{u}lich, 52425~J\"{u}lich, Germany}

\date{\today}
\vspace{1cm}

\begin{abstract}
We propose to tailor exchange interactions in magnetic monolayer
films by tuning the adjacent non-magnetic substrate. 
As an
example, we demonstrate a ferromagnetic-antiferromagnetic phase
transition for one monolayer Fe on a Ta$_{x}$W$_{1-x}$(001)
surface as a function of the Ta concentration. At the critical Ta
concentration, the nearest-neighbor exchange interaction is small 
and the magnetic phase space is dramatically broadened. 
Complex magnetic order such as spin-spirals, multiple-$Q$,
or even disordered local moment states can occur, offering
the possibility to store information in terms of ferromagnetic dots in an 
otherwise zero-magnetization state matrix.
\end{abstract}

\maketitle

%%%%%%%%%%%%%%%%%%%%%%%%   INTRODUCTION    %%%%%%%%%%%%%%%%%%%%%%%%

Magnetic systems play a central role in today's information
technology and our ability to control and tailor their properties
may open new vistas to future device
concepts. Materials structured on a nanometer scale such as
atomically thin films proved to be a rich field for novel
magnetic properties.
While our understanding of magnetic systems has
tremendously increased over the past 25 years,
to control magnetic order in a specific system and tailor
materials with desired magnetic properties remains the grand
challenge of research in magnetism.

So far the attempts to tune the magnetic state of surfaces and
ultra-thin films have focussed on alloy formation where the 
concentration of the magnetic components is altered to 
optimize the magnetic 
properties~\cite{Turek98PRB, Ponomareva03PRB, Offi02PRB,Kuch04PRL, Kuch02PRB,Skubic06PRL}.
In this letter, we propose a completely different route. Based on the 
surprising observation of an
antiferromagnetic (AFM) order of one monolayer (ML) Fe on W(001) and
the prediction of 
the ferromagnetic (FM) order on Ta(001)~\cite{KFB2005}, we suggest to
tune magnetic interactions in ultra-thin films by modifying only
the band-filling of the substrate, via the formation of a Ta-W alloy,
without altering or diluting  the magnetic monolayer itself.

We employ first-principles calculations to show that the
nearest-neighbor exchange interaction in one ML Fe on the
(001) surface of a Ta$_x$W$_{1-x}$ alloy can be continuously tuned
from FM to AFM coupling by varying the Ta concentration $x$.
At the substrate composition of small nearest-neighbor exchange
interaction, we find that 
higher order spin interactions beyond
the Heisenberg model, such as biquadratic or four-spin interactions,
may stabilize complex non-collinear magnetic structures.
In this case, we also consider the role of chemical disorder
which might prevent any stable magnetic order due to the small
energy scale involved and lead for example to a spin-glass.
The substrate turns out to be a tuner also for 
a magnetic order-disorder phase transition.
At the corresponding Ta concentration, a highly frustrated material is formed and 
can be used to store information in the form of FM dots in a zero-magnetization state matrix,
opening the way to a new class of material for magnetic storage devices.

While we consider a single model 
system, which allows to
isolate magnetic effects from structural and chemical ones,
depending on surface orientation, substrate element, and overlayer,
a rich variety of similar systems are possible.
For example, metallic magnets with small 
exchange coupling have been recently reported \cite{HunterDunn05PRL,vonbergmann06PRL}.

%%%%%%%%%%%%%%%%%%    Calculational details    %%%%%%%%%%%%%%%%%%%%%%%

We have determined the electronic and magnetic properties of one
ML Fe on the (001) surface of \wta by performing
density-functional theory calculations 
in the generalized-gradient approximation to the 
exchange-correlation functional \cite{ZY1998}.
The substitutional \wta  random alloy has been modelled in the spirit of 
the virtual crystal
approximation (VCA)~\cite{Turchi01PRB} by a substrate of fictitious atoms
with fractional atomic numbers related to the Ta composition $x$,
ranging linearly between 73 (Ta) and 74 (W). 
The corresponding fractional electronic 
charge preserves charge neutrality and accounts for the 
variation of the band-filling  originating from alloying. 
Vegard's law was adopted to interpolate between the lattice constants
of Ta ($3.301$ \AA) and W ($3.165$ \AA). 
A fixed surface relaxation of 18\% was assumed, corresponding to the
relaxation of the FM Fe monolayer on pure W(001).
Based on additional studies on the effect of the lattice constant on the 
magnetic properties, the tiny deviations from Vegard's law~\cite{Turchi01PRB} 
and the deviation of the relaxation from the average value~\cite{NOTE3} can be safely 
neglected.
The calculations have been carried out with the full-potential linearized
augmented plane wave (FLAPW) method in film geometry, 
as implemented in the {\sc fleur} code \cite{FLEUR}. 
Spin spirals have been computed in the $p(1\times1)$
unit cell, exploiting the generalized Bloch theorem \cite{Kurz04PRB}.
The computational parameter were chosen according to Ref.~\cite{KFB2005}.

\begin{figure}
\begin{center}
\includegraphics[scale=0.6,angle=0]{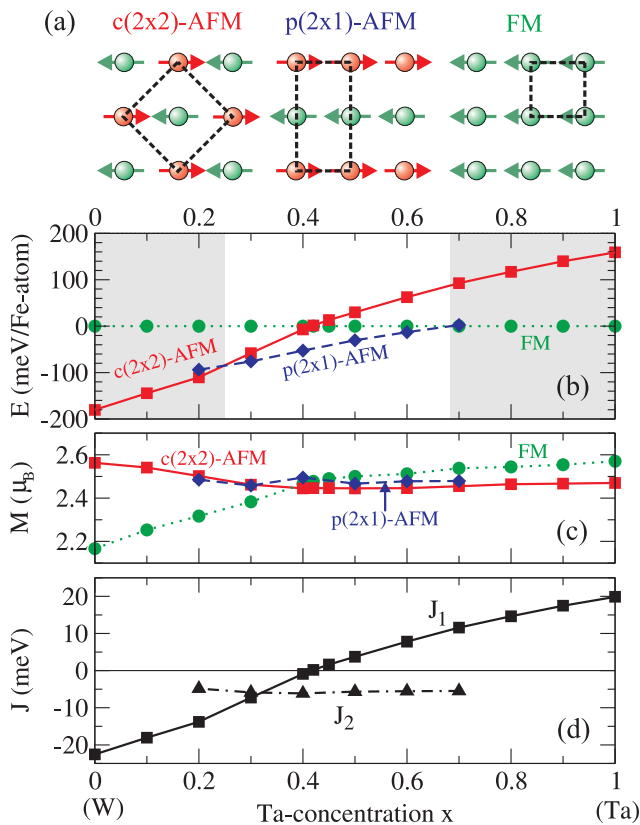}
\caption{(color online) (a) Unit cell sketch of the investigated magnetic configurations.
(b) Total energy, (c) magnetic moment, and (d) exchange constants for 1 ML Fe/\wta\, 
as calculated by the FLAPW method.}
\label{Fig1}
\end{center}
\vspace{-.5cm}
\end{figure}

%%%%%%%%%%%%%%%%%%%%%%%%   BODY    %%%%%%%%%%%%%%%%%%%%%%%
W and Ta are adjacent elements of the periodic table with similar
properties. Both crystallize in the bcc structure with
comparable lattice constants.
W has one $d$ electron more than Ta.
While one ML Fe exhibits a \cA\, state, 
Fig.~\ref{Fig1}(a), if grown on W(001), it is FM
on Ta(001)~\cite{KFB2005}. Hence, the magnetic configuration in the Fe
layer is related to the substrate $d$-band filling that
affects the position of the substrate $d$-band relative to the Fermi energy (E$_\mathrm{F}$)
and controls by hybridization the position of the Fe states at
E$_\mathrm{F}$.
For $3d$ metals with large magnetic moments it is well-established 
that in good approximation the magnetic
configuration with the lowest minority density of states at E$_\mathrm{F}$ 
exhibits the lowest
energy (e.g.\ compare  Fig.~3 and 4 in Ref.~\onlinecite{FHBB2005}).
By forming \wta alloys with different compositions we can adjust 
the $5d$-band filling or the position of the minority Fe $d$-state at E$_\mathrm{F}$
and continuously investigate
the transition between the AFM and FM state of the Fe overlayer in Fe/W(001) and
Fe/Ta(001), respectively. This is 
expected to be experimentally feasible due to the full 
miscibility of W and Ta for every concentration \cite{Massalski86}.

This scenario is presented in Fig.~\ref{Fig1}. We initially focus
on three collinear configurations, Fig.~\ref{Fig1}(a): \cA, 
\pA, also referred to as row-wise AFM (RW-AFM), and FM.
The magnetic ground-state
is obtained from the
total energy, Fig.~\ref{Fig1}(b), as a function of the Ta
concentration, $x$, relative to the FM state.
At small Ta concentrations up to about 25\%, the Fe monolayer
exhibits a \cA\, order, while at large $x$, roughly beyond 70\%,
the ground state is FM.
In the intermediate range the RW-AFM state is
energetically favorable~\cite{NOTE1}.

In order to analyze these {\em ab-initio} calculations in terms of
exchange interactions, we can map the results onto a classical
Heisenberg Hamiltonian, $ H=-\sum_{i<j}J_{ij} \mathbf{s}_i \cdot
\mathbf{s}_j$, where $J_{ij}$ is the exchange interaction between
spins at lattice sites $i$ and $j$ pointing in the direction of
the unit vectors $\mathbf{s}_i$ and $\mathbf{s}_j$, respectively.
Such a model with fixed spin values is legitimated by the 
weak dependence of the magnetic moment on the Ta concentration
and magnetic order, Fig.~\ref{Fig1}(c). For a monolayer on a square lattice
the nearest neighbor (nN) and next-nearest neighbor (nnN) exchange
constants $J_1$ and $J_2$, respectively, can be extracted from the
energies of the FM, $c(2\times2)$-, and \pA\, states.

The nN exchange constant varies as a
function of the Ta concentration,
Fig.~\ref{Fig1}(d), and mimics the energy of the \cA~configuration,
Fig.~\ref{Fig1}(b).
In contrast, we observe only a weak dependence
for the nnN coupling which is on the order of 
$J_2\approx-5$ meV.
The strength of the typically dominating 
nN exchange interaction in a magnetic film can therefore be tuned
by selecting the proper substrate composition, without requiring
structural or chemical modifications
of the film. This opens the way to a dramatically broader 
magnetic phase space, as we will see in the following.

\begin{figure}
\begin{center}
\includegraphics[scale=0.48,angle=0]{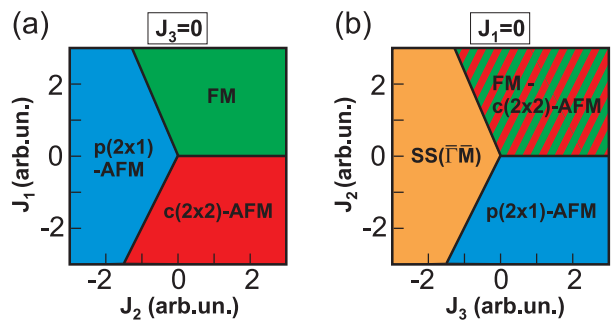}
\caption{(color online) 
Phase diagrams for
2D systems on a square lattice based on the Heisenberg
model. (a) If $J_1$ and $J_2$ dominate, only collinear states are
possible. (b) If $J_1$ is small compared to $J_3$, the FM and the
\cA\, configurations are degenerate and 
spin spirals with $\mathbf{q}$ along $\overline{\Gamma}-\overline{\text{M}}$
can occur.
} \label{Phasediag}
\end{center}
\vspace{-.1cm}
\end{figure}

The possible magnetic states in a two-dimensional (2D) film on a square lattice can
be studied in the magnetic phase diagrams derived within the
Heisenberg model, Fig.~\ref{Phasediag}. For dominating nN exchange
interaction only two magnetic states can occur, either the FM ($J_1>0$) state
or the \cA\ one ($J_1<0$),
 Fig.~\ref{Phasediag}(a). Even in the presence of nnN exchange the
magnetic states are still limited to collinear solutions.
However, if $J_1$ is negligible or at least comparable to
interactions
beyond second nearest neighbors, complex non-collinear magnetic
phases appear, Fig.~\ref{Phasediag}(b).
For example, a considerable value of $J_3$
leads to so-called spin-spiral states. Flat spin spirals are the
general solution of the Heisenberg Hamiltonian and are
characterized by a
wave vector $\mathbf{q}$. For a given
$\mathbf{q}$, the magnetic moment of an atom at site $\mathbf{R} $
points in the direction $\hat{\mathbf{s}}\left( \mathbf{R}\right)
= \left(
           \cos \left( \mathbf{q} \cdot \mathbf{R} \right),
           \sin \left( \mathbf{q} \cdot \mathbf{R} \right),
           0
\right)$.

\begin{figure}
\begin{center}
\includegraphics[scale=0.64,angle=-0]{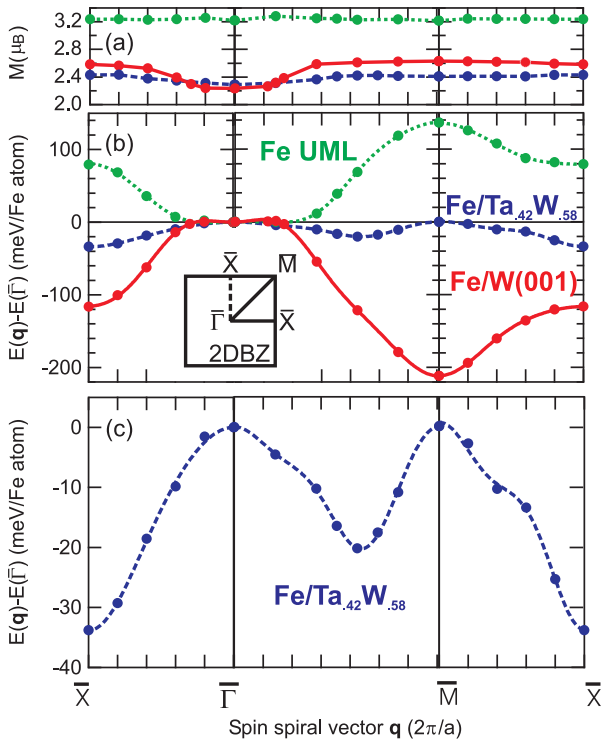}
\caption{(color online) (a) Magnetic moment and (b) energy dispersion for spin
spirals of an Fe UML at the W(001)
lattice constant (dotted green line), 1 ML Fe/W(001) (solid red
line), and 1 ML Fe/\wtax(001) (dashed blue line). 
(c) Dispersion for 1 ML Fe/\wtax(001) on a
larger scale. Symbols denote {\em ab-initio} calculations
and lines are Heisenberg fits up to eight neighbors.
}
\label{Spinspirals}
\end{center}
\vspace{-.5cm}
\end{figure}
Based on these phase diagrams, the Ta concentration
of very small $J_1$, i.e.\ close to $x=42\%$, is
particularly intriguing, as interactions beyond nN
will determine the magnetic ground state and non-collinear
solutions are likely 
\cite{HunterDunn05PRL,vonbergmann06PRL}.
This suggests to extend
the calculations by including spin spirals for vectors $\mathbf{q}$
along the high-symmetry lines of the irreducible 2D Brillouin zone. 
The high symmetry points represent the
previously discussed collinear states: $\overline{\Gamma}$,
$\overline{\text{M}}$, and $\overline{\text{X}}$
correspond to the FM, \cA, and \pA\, state, respectively.

In order to emphasize the interplay of overlayer-substrate
interaction and magnetic order, we study the energy dispersion
$E(\mathbf{q})$ for one ML Fe in different environments,
 Fig.~\ref{Spinspirals}. The unsupported monolayer (UML) of Fe(001) 
at the W lattice constant is an example
of a system dominated by $J_1$, i.e.\ the FM state is the
ground-state, the \cA\ state is unstable, and the RW-AFM state is
metastable, with large energy differences among them.
From a fit based on the Heisenberg model, we obtain the exchange
constants and find $J_1=17$ ~meV and $J_1 \gg J_i$, for $i>1$, in
good agreement with the frozen magnon calculation by Sandratskii
\emph{et al.} \cite{Sandratskii06PRB}. For 1 ML Fe/W(001) the
hybridization with the substrate modifies the electronic structure.
The dispersion is reversed, with the
minimum at the $\overline{\text{M}}$ point. This implies that $J_1$ is
again the leading term, but with a negative sign: $J_1=-26$ ~meV.

However, for 1 ML Fe/\wtax (001) the dispersion is strikingly
different from the two previous cases.
Since $J_1\approx0$ for this particular substrate composition, the
FM and \cA\, states are degenerate and the energy scale is
strongly reduced, Fig.~\ref{Spinspirals}(b). 
Interestingly, a
metastable spin-spiral state is found along the
$\overline{\Gamma}$-$\overline{\text{M}}$ direction, 20 meV/Fe-atom lower
than the FM one, Fig.~\ref{Spinspirals}(c). The RW-AFM state
is the global
energy minimum. The fit of the dispersion reveals that the leading
interaction is $J_2= -6$~meV and the system is in the lower left
part of the $p(2\times 1)$ region in the phase diagram of Fig.~\ref{Phasediag}(b).

 \begin{figure}[!t]
 \begin{center}
 \includegraphics[scale=0.35,angle=0]{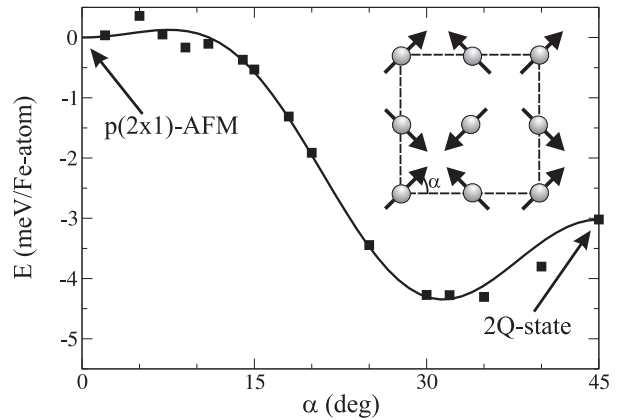}
 \caption{Total energy of the
 configuration depicted in the inset for 1 ML Fe/\wtax(001), 
 as a function of the 
 angle $\alpha$.
 }
 \label{Eofalpha}
 \end{center}
 \vspace{-.5cm}
 \end{figure}
Due to the surface symmetry, 
there are two equivalent $\overline{\text{X}}$ states in the
Brillouin zone (see inset of Fig. \ref{Spinspirals}(b)), corresponding to two
degenerate \pA\, configurations with perpendicular orientations of
the ferromagnetically coupled rows. 
Any superposition of these two spin spirals, a so-called multi-$Q$
state, is a degenerate solution of the Heisenberg Hamiltonian.
However, the degeneracy with the \pA\, state can be lifted by
higher order interactions beyond the Heisenberg model, such as
the four-spin and the biquadratic ones~\cite{Kurz01PRL}.

Such interactions are implicitly included in the exchange
correlation potential, and we can evaluate their magnitude from
first principles.
We performed
calculations in the $p(2\times2)$ unit cell rotating the moments
on all sites by an angle $\alpha$ as depicted in the inset of
 Fig.~\ref{Eofalpha}. $\alpha=0^\circ$ corresponds to the
RW-AFM state while for $\alpha=45^\circ$ we obtain the 2$Q$-state,
a 2D non-collinear structure with perpendicular adjacent moments.
Since all these states are degenerate within the Heisenberg
model, the total energy difference depends only on higher order
interactions.
We find that non-collinear states gain energy on the order of 5 meV 
due to these interactions, the
minimum being at $\alpha=31^\circ$.
The fitting revealed that even terms 
beyond the biquadratic and four-spin interactions
are present in this system.
The moment arrangement can be slightly modified by the magnetocrystalline anisotropy.
E.g., an out-of-plane easy axis with an anisotropy energy on the order of 2 meV, 
similar to that of 1 ML Fe/W(001) \cite{KFB2005}, would decrease $\alpha$ from $31^\circ$ to
about $20^\circ$, based on Fig.~\ref{Eofalpha}, but would not prevent the non-collinear order.

%-----------------------------------------------------------------------

 \begin{figure}[!t]
 \begin{center}
 \includegraphics[scale=0.47,angle=0]{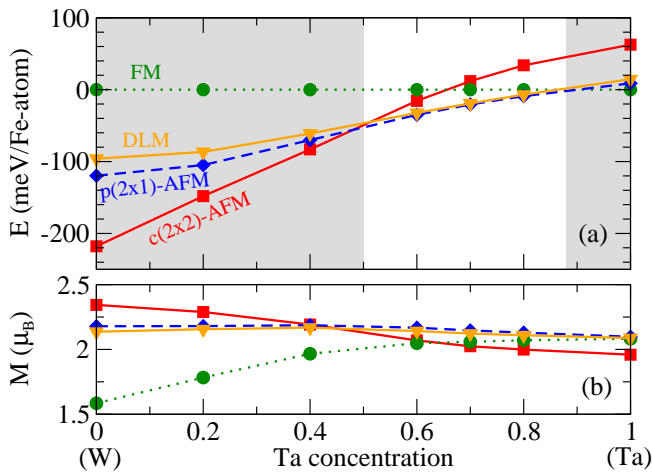}
 \caption{(color online) (a) Total energy and (b) magnetic moment for 1 ML Fe/\wta (001)
 as a function of the Ta concentration $x$, as obtained with the TB-LMTO method.
 }
 \label{CPA}
 \end{center}
 \vspace{-.5cm}
 \end{figure}
The small energy scale found in the spin spiral as well as in the
2$Q$-state calculation for 1 ML Fe/\wtax (001) indicates that only a
small amount of energy is required to rotate the magnetic moments and that
there is a competition between several magnetic interactions.
This suggests a frustrated system.  Under these circumstances, the
disordered local moment (DLM) state with local moments pointing in
random directions and zero net magnetization needs
to be considered.
We have evaluated its energy using the 
tight-binding linear muffin-tin orbital (TB-LMTO) method
in the atomic sphere approximation within the framework of the
coherent-potential approximation (CPA) \cite{Gyorffy85JPF,Turek97}.
Concerning the magnetically ordered states, we found that for small, intermediate, 
and large Ta concentrations the ground state is \cA, \pA, and FM, 
respectively (Fig.~\ref{CPA}).
The qualitative agreement between FLAPW-VCA and TB-LMTO-CPA
 proves that the VCA is a good approximation
to the CPA for the treatment of the alloy. 
Interestingly, for intermediate compositions the DLM and the \pA\,
states are degenerate, within the computational accuracy. 
Note that the DLM state is often encountered as a spin-glass-like
ground state in disordered bulk alloys,
such as fcc Ni$_{0.80}$Mn$_{0.20}$ \cite{Akai93PRB}.
All important conditions for the spin-glass (SG) arrangement,
namely competing ferro- and antiferromagnetic exchange interactions
accompanied by chemical and/or topological disorder \cite{Moorjani84},
are fulfilled in the present case owing to the random Ta-W substrate.
At around $x=85$\% Ta concentration, the SG, FM, and \pA\, states
are degenerate. 
This offers the great perspective of imprinting
magnetic information as nanoscale dots. 
E.g., cooling down the dot, after local heating at elevated temperatures,
with or without an external magnetic field allows to systematically select 
the FM or the zero-magnetization SG state.

%%%%%%%%%%%%%%%%%%%%%   SUMMARY   %%%%%%%%%%%%%%%%%%%%%%%%%%%%%
In summary, we introduced
here a new model system to study magnetic phase transitions
in low-dimensional magnets allowing to sweep through a series of 
magnetic phases without changing the structure or chemical composition
of the magnet. The system offers the possibility to store information
in terms of FM dots in an otherwise zero-magnetization state matrix.
%%%%%%%%%%%%%%%%%%  ACKNOWLEDGEMENT  %%%%%%%%%%%%%%%%%%%%%%%%%%

S.B. benefitted from discussions with F.~F\"orster.
S.H. thanks the Stifterverband f\"ur die Deutsche
Wissenschaft and the Interdisciplinary Nanoscience Center Hamburg
for financial support.
The work of I.T. was partly supported by the Grant Agency of the
Academy of Sciences of the Czech Republic (Project No. A100100616).

%%%%%%%%%%%%%%%%%%   BIBLIOGRAPHY   %%%%%%%%%%%%%%%%%%%%%%%%%%%%%
\vspace{-.7cm}
%\bibliography{MyReferenceList}

\begin{thebibliography}{21}
\bibitem[{NOT({\natexlab{a}})}]{NOTE1}
\eprint{The energy difference between the FM and the \cA\, state has been
  calculated for the particular case of $x=50\%$ also with an ordered alloyed
  substrate, composed by 50\% W and 50\% Ta atoms in a $c(2\times2)$
  arrangement. A discrepancy of only 7.3 meV/Fe-atom was found with respect to
  the uniform slab with an atomic number of 73.5. This validates to neglect the
  chemical inhomogeneity of the substrate for a calculation of $J_1$ and
  supports the idea that the key effect of W-Ta alloying is a modification of
  the $5d$ band filling.}
\end{thebibliography}

\end{document}